\documentclass[fleqn,twoside]{article}

\usepackage{espcrc2}

\usepackage{graphicx}
\usepackage{epsfig}

\usepackage[figuresright]{rotating}

\newcommand{\AmS}{{\protect\the\textfont2
A\kern-.1667em\lower.5ex\hbox{M}\kern-.125emS}}

\def\etal{{\it et al}}

\title{Construction, Pattern Recognition and Performance of the CLEO III LiF-TEA RICH Detector}
\author{M. Artuso, R. Ayad, K. Bukin, A. Efimov, C. Boulahouache,
E. Dambasuren, S. Kopp, R. Mountain, G. Majumder, S. Schuh,
T. Skwarnicki, S. Stone, G. Viehhauser, J. C. Wang, and
\address{Syracuse University,  Syracuse, NY 13244-1130, U. S. A.}\\
T. E. Coan, V. Fadeyev, Y. Maravin, I. Volobouev, J. Ye, and
\address{Southern Methodist University, Dallas, TX 75275-0175}\\
S. Anderson, Y. Kubota, and A. Smith
\address{University of Minnesota, Minneapolis, MN 55455-0112}\\}
\begin{document}

\begin{abstract}

We briefly describe the design, construction and performance of
the LiF-Tea RICH detector built to identify charged particles in
the CLEO III experiment. Excellent $\pi/K$ separation is
demonstrated.

\vspace{1pc}

\end{abstract}


\maketitle

\section{INTRODUCTION}

\subsection{The CLEO III Detector}

The CLEO III detector was designed to study decays of $b$ and $c$ quarks,
$\tau$ leptons and $\Upsilon$ mesons produced in $e^+e^-$ collisions
near 10 GeV center-of-mass energy. The new detector is an upgraded
version of CLEO II \cite{CLEOII}. It contains a new four-layer silicon strip
vertex detector, a new wire drift chamber and a particle identification system
based on the detection of Cherenkov ring images. Information about CLEO III
is available elsewhere \cite{Artu98,Kopp96}.

CLEO II produced many physics results, but was hampered by its limited
charged-hadron identification capabilities.
Design choices for particle identification were
limited by radial space and the necessity of minimizing the material in front
of the CsI crystal calorimeter. The CsI imposed a hard outer radial limit and the
desire for maintaining excellent charged particle tracking imposed a
lower limit, since at high momentum the error in momentum is
proportional to the square of the track length. The particle identification
system was allocated only 20 cm of radial space, and this limited the
technology choices. We were also allowed a total material thickness corresponding to only 12\% of a radiation length.

\section{DETECTOR DESCRIPTION}

\subsection{Detector Elements}

\begin{figure}[htb]
\vspace{-.5cm}
\includegraphics[height=2.4in]{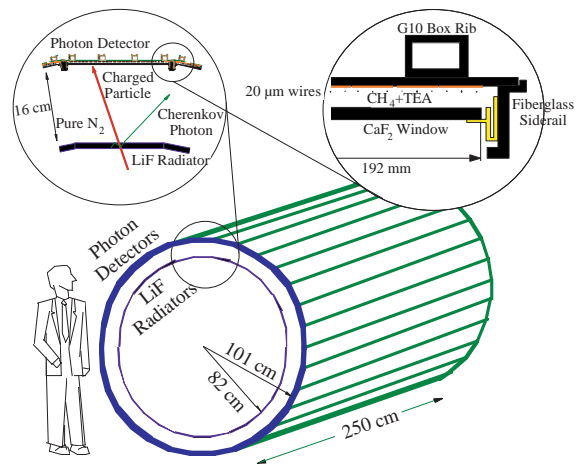}
\vspace{-1cm} \caption{\label{RICH_descrip}Outline of CLEO III
RICH design.} \vspace{-0.7cm}\end{figure}

The severe radial spatial requirement forces a thin, few cm
detector for Cherenkov photons and a thin radiator. Otherwise the
photons have little distance to travel and it becomes very
difficult to precisely measure the photon angles. In fact, the
only thin photon-detectors possible in our situation were wire
chamber based, either CsI or a mixture of triethylamine (TEA) and
methane. Use of CsI would have allowed us to use a liquid freon
radiator with quartz windows in the system using the optical
wavelength region from about 160-200 nm. However, at the time of
decision, the use of CsI was far from proven and, in any case,
would have imposed severe constraints on the construction process
which would have been both difficult and expensive. Thus we chose
TEA + CH$_4$ and used Cherenkov photons between 135-165 nm
generated in a 1 cm thick LiF crystal and used CaF$_2$ windows on
our wire chambers (LiF windows were used on 10\% of the chambers)\cite{Arno92}.

Details of the design of the CLEO III RICH have been discussed before
\cite{testbeam}. Here we briefly review the main elements.
Cherenkov photons are produced in a LiF radiator. The photons then enter
a free space, an ``expansion volume," where the cone of Cherenkov light
expands. Finally the photons enter a detector consisting of
multi-wire proportional chambers filled with a mixture of TEA
and CH$_4$ gases.
No light focusing is used; this is called ``proximity-focusing''
\cite{t+j}. The scheme is shown in the upper left of Fig.~\ref{RICH_descrip}.

There are 30 photon detectors around the cylinder. They subtend the same azimuthal angle as the radiators, which are also segmented into 14 sections along their length of the cylinder. The gap between the radiators and detectors, called the ``expansion gap", is filled with pure N$_2$ gas. The wire chamber design is shown in Fig.~\ref{RICH_descrip}.

\subsection{Radiators}

\begin{figure}[htb]
\vspace{-.3cm}
\centerline{\epsfig{figure=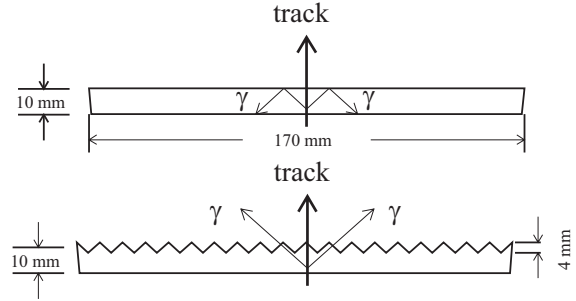,height=1.6in}}
\vspace{-.6cm} \caption{\label{radiators}Sketch of a plane
radiator (top) and a sawtooth radiator (bottom). Light paths
radiated from a charged track normal to each radiator are shown.}
\vspace{-0.7cm}\end{figure}

LiF was chosen over CaF$_2$ or MgF$_2$, both of which are
transparent in the useful wavelength region, because of smaller
chromatic error. Originally all the radiators were planned to be 1
cm thick planar pieces. However, since the refractive index of LiF
at 150 nm is 1.5, all the Cherenkov light from tracks normal to
the LiF would be totally internally reflected as shown in
Fig.~\ref{radiators} (top). We could have used these flat
radiators, but we would have had to tilt them at about a
15$^{\circ}$ angle. Instead we developed radiators with striations
in the top surface, called ``sawtooth" radiators \cite{efimov}, as
shown in Fig.~\ref{radiators} (bottom).

The overall radiator shape approximates a cylinder of radius 82
cm. Individual radiators are placed in 14 coaxial rings of 30
crystals each, centered around the beam line and symmetrically
positioned about the interaction point. Only the inner four radiators are sawtooths.(The 30 crystals segments
are parallel to the wire chambers.) Inter-crystal gaps are
typically 50 $\mu$m. The crystals are attached to the exterior
surface of a 1.5 mm thick carbon fiber shell with a low outgassing
epoxy.

\subsection{Photon Detectors}
Construction was carried out in a class 100 clean room that was dehumidified below 35\%. Granite tables were used that were flat over the entire surface of a photon detector module to better than 15 $\mu$m.

The photon detectors have segmented cathode pads 7.5 mm (length) x 8.0 mm (width) etched onto G10 boards. The pad array was formed from four individual boards, with 24 x 80 pads, with the latter separated into two 40 pad sections with a 6 mm gap. Each board was individually flattened in an oven and then they were glued together longitudinally on a granite table where reinforcing G10 ribs were also glued on. There are 4 longitudinal ribs that have a box structure. Smaller cross ribs are placed every 12 cm for extra stiffening. The total length was 2.4 m.

Wire planes were separately strung with 20 $\mu$m diameter gold
plated tungsten with a 3\% admixture of rhenium produced by LUMA; the wire pitch
was 2.66 mm, for a total of 72 wires per chamber. The wires were
placed on and subsequently glued to precision ceramic spacers 1 mm
above the cathodes and 3.5 mm to the CaF$_2$ windows, every 30 cm.
We achieved a tolerance of 50 $\mu$m on the wire to cathode
distance. The spacers had slots in the center for the glue bead.

Eight 30 cm x 19 cm CaF$_2$ windows were glued together in precision jigs lengthwise to form a 2.4 m long window.
Positive high voltage is applied to the anode wires, while - high voltage is put on 100 $\mu$m wide silver traces deposited on the CaF$_2$. The spacing between the traces is 2.5 mm. To maintain the ability of disconnecting any faulty part of a chamber, the wire HV is distributed independently to 3 groups of 24 wires and the windows are each powered separately.

\subsection{Electronics}

\begin{figure} [htb]
\vspace{-.01cm} \centerline{\epsfxsize 2.5in
\epsffile{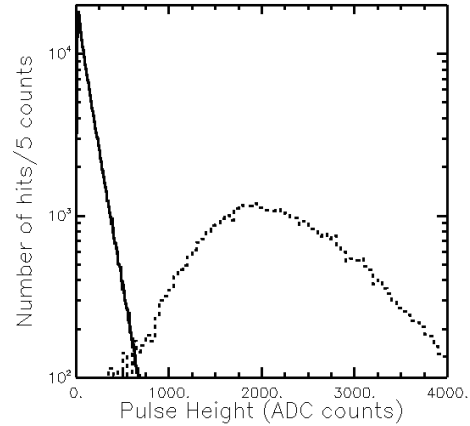}} \vspace{-.7cm}
\caption{\label{charge_distribution} Pulse height distributions
from pad clusters containing single photons (solid histogram) and
charged tracks (dashed histogram). The line shows a fit of photon
data to an exponential distribution. One ADC count corresponds to
approximately 200 electrons. The charged track distribution is
affected by electronic saturation. The wire voltage is +1.5 kV and the window voltage -1.2 kV.}
\vspace{-0.7cm}
\end{figure}

The position of Cherenkov photons is measured by sensing the
induced charge on array of 7.5 mm x 8.0 mm cathode pads. Since the
pulse height distribution from single photons is expected to be
exponential at low to moderate gas gains \cite{expon}, this requires the use of low noise
electronics. Pad clusters in the detector can be formed from
single Cherenkov photons, overlaps of more than one Cherenkov
photon or charged tracks. In Fig.~\ref{charge_distribution} we
show the pulse height distribution for single photons, and charged
tracks. We can distinguish somewhat between single photons hitting
the pad array and two photons because of the pulse height shapes
on adjacent pads.
  The charged tracks give very large pulse heights because they are traversing
4.5 mm of the CH$_4$-TEA mixture. 

To have as low noise electronics as possible, a dedicated VLSI
chip, called VA\_RICH, based on a very successful chip developed
for solid state applications, has been designed and produced for
our application at IDE AS, Norway. We have fully characterized
3,600 64 channel chips, mounted on hybrid circuits. For moderate
values of the input capacitance $C_{in}$, the equivalent noise
charge measured $ENC$ is found to be about:
\begin{equation} ENC = 130 e^- + (9 e^-{\rm /pF}) \times C_{in}~~. \end{equation}
Its dynamic range is between 450,000 and 900,000 electrons, depending upon
whether we choose a bias point for the output buffer suitable for signals of
positive or negative polarity or we shift this bias point to have the maximum
dynamic range for signals of a single polarity.

In our readout scheme we group 10 chips in a single readout cell communicating
with data boards  located in VME crates just outside the detector
cylinder. Chips in the same readout cell share the same cable, which routes
control signals and bias voltages
 from the data boards and output signals to the data boards.
  Two VA\_RICH chips are mounted using wire bonds on one hybrid circuit that
is  attached via two miniature connectors to  the back of the cathode board
of the photon detector.

The analog output of the VA\_RICH is transmitted to the data boards as a
differential current, transformed  into a voltage by transimpedance amplifiers
and digitized by a 12 bit differential ADC. These receivers are  part of  very
complex data boards which perform several important analog and digital functions.
Each board  contains 15 digitization circuits and three analog power supply
sections providing the voltages and currents to  bias the chips, and
calibration circuitry. The digital component of these boards contains a
sparsification  circuit, an event buffer, memory to store the pedestal values,
and the interface to the VME cpu.

Coherent noise is present. We eliminate this by measuring the
pulse heights on all the channels and performing an average of the
non-struck channels before the data sparsification step
\cite{DSP}. The pedestal width (rms) changes from 3.6 to 2.5
channels with and without this coherent noise subtraction,
respectively. The total noise of the system then is $\sim$500
electrons rms.

\subsection{Gas System}
The gas system supplies several distinct volumes. The systems
must: supply CH$_4$-TEA to 30 separate chambers, supply
super-clean N$_2$  to the expansion gap, supply super-clean N$_2$
to a sealed single volume surrounding all the chambers, called the
electronics volume, since this is the region where the front-end
hybrid boards are present. In addition we need to test CH$_4$-TEA
for the ability to detect photons and test the output N$_2$ for
purity.

It is of primary importance that the gas system must NOT destroy any of the thin CaF$_2$ windows. We use computerized pressure and flow sensors with PLC controllers. The gas system works great. N$_2$  transparency is
$>$99\%. Nothing has been broken!

\section{OPERATING EXPERIENCE}

The detector has been in operation since September of 1999.
All but $\sim$2\% of the detector is functioning. We lost 1\% due to the
breaking of one wire after about one year of operation.
We have also lost 2\% of the electronics chips.

\section{OFF-LINE DATA ANALYSIS AND PHYSICS PERFORMANCE}

\subsection{Noise filtering}

Coherent noise suppression and data sparsification are performed
on-line to eliminate the Gaussian part of the electric noise. A
small non-Gaussian component of the coherent electric noise is
eliminated off-line, by using an algorithm too complicated for use
in the data board DSP. The incoherent part of non-Gaussian noise
was eliminated by off-line pulse height thresholds adjusted to
keep occupancy of each channel below 1\%. Finally we eliminate
clusters of cathode pad hits that are extended along the anode
wires, but are only 1-2 pads wide in the other direction.

\subsection{Cherenkov Images}

\begin{figure} [htb]
\vspace{-.8cm} \centerline{\epsfysize
1.38in\epsffile{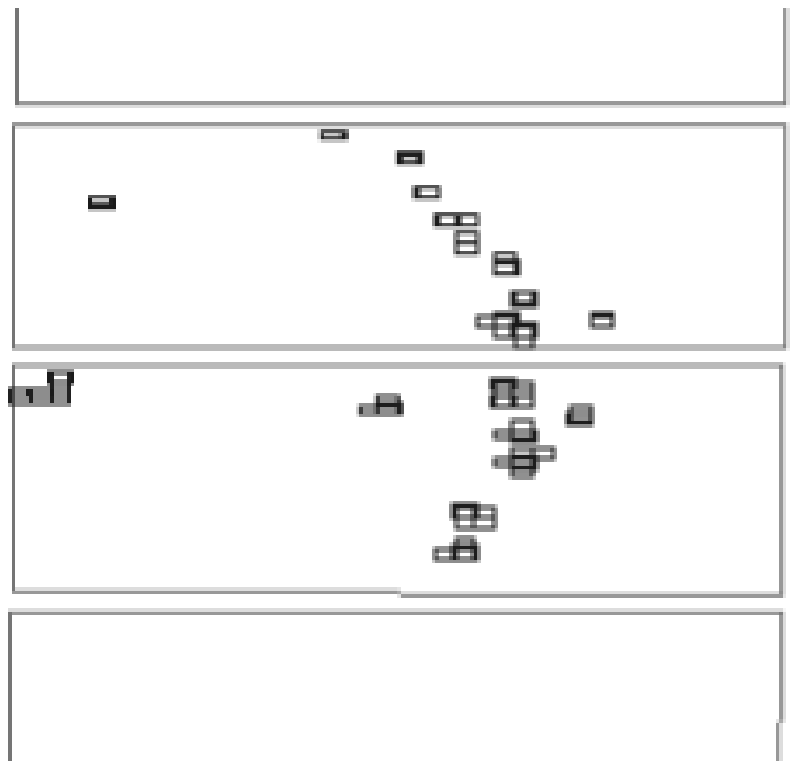}
            \epsfysize 1.43in\epsffile{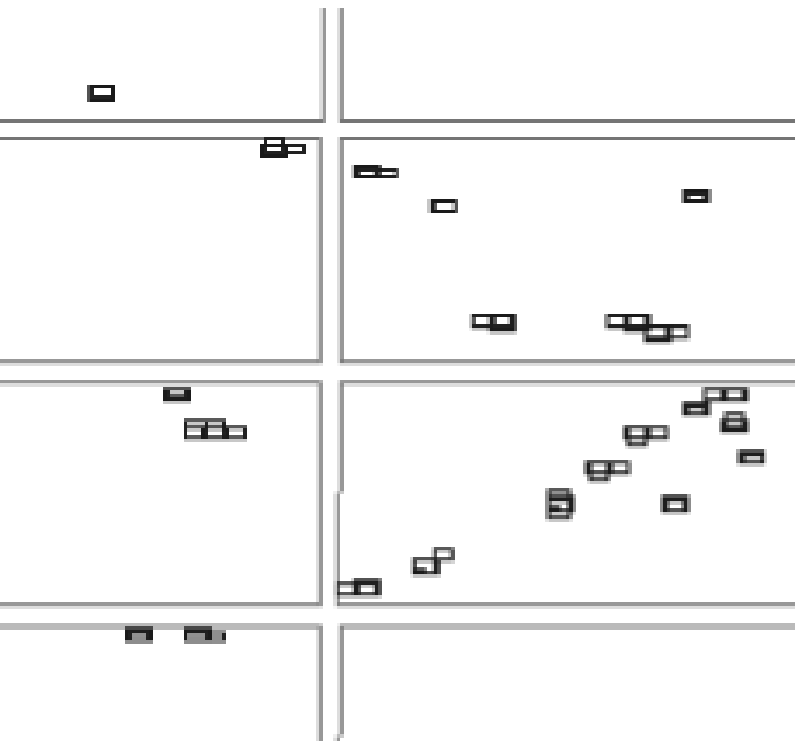}}
\vspace{-.7cm}
\caption{\label{images}
      Hit patterns produced by the particle passing through the plane
      (left) and sawtooth (right) radiators.}
\vspace{-0.7cm}\end{figure}

We show in Fig.~\ref{images} the hit pattern in the detector for a Bhabha
scattering event ($e^+e^-\to e^+e^-$) for track entering the
plane (left image) and
sawtooth (right image) radiators.
The shapes of the Cherenkov ``rings" are different in the two cases,
resulting from refraction when leaving the LiF radiators.
The hits in the centers of the images are produced by
the electron passing the RICH MWPC.

\subsection{Clustering of Hits}

The entire detector contains 230,400 cathode pads, which are
segmented into 240 modules of $24\times48$ pads separated by the
mounting rails and anode wire spacers. We cluster pad hits in each
module separately. Pad hits touching each other either by a side
or a corner form a ``connected region." Each charged track
reconstructed in the CLEO-III tracking system \cite{Peterson} is
projected onto the RICH MWPC and matched to the closest connected
region. If the matching distance between the track projection and
the connected region center is reasonably small and the total
pulse height of the connected region sufficiently high we
associate this group of hits with the track. Local pulse height
maxima in the remaining connected regions, so called ``bumps"  are
taken as seeds for Cherenkov photons. We allow the pulse height
maxima to touch each other by corners if the pulse height in the
two neighboring pads is small relative to both bump hits. Hits
adjacent to the bumps on the sides are assigned to them in order
of decreasing bump pulse height.

To estimate the position of the photon conversion point we use the
center-of-gravity method corrected for the bias towards the
central pad. For many Cherenkov photons we are able to detect
induced charge in only one pad. Since the pad dimensions are about
$8\times8$ mm$^2$, the position resolution in this case is
$8/\sqrt{12}=2.3$ mm. For charged track intersections, which
induce significant charge in many pads, the position resolution is
$0.76$ mm. The position resolution for Cherenkov photons which
generate multiple pad hits is somewhere in between these two
values. In any case, the photon position error is not a
significant contribution to the Cherenkov angle resolution (see
below).

\subsection{Corrections to the Track Direction}

The resolution of the CLEO-III tracking system is very good in the
bending view (the magnetic field is solenoidal in CLEO)
\cite{Peterson} . The track position and inclination angle along
the beam axis is measured less precisely, with the silicon vertex
detector playing the dominant role. The rms of the observed RICH
hit residual is 1.7 mm. Since the RICH hit position resolution is
0.76 mm as measured by the residual in the perpendicular
direction, the RICH MWPC can clearly help in pinning down the
track trajectory. This, in turn, improves Cherenkov resolution,
especially for the flat radiators for which we observe only half
of the Cherenkov image and thus are quite sensitive to the
tracking error. The improvement is as much as 50\%\ is some parts
of the detector.

\subsection{Reconstruction of Cherenkov Angle}

Given the measured position of the Cherenkov photon conversion
point in the RICH MWPC, the charged track direction and its
intersection point with the LiF radiator, we calculate a Cherenkov
angle for each photon-track combination ($\theta_\gamma$). We use
the formalism outlined by Ypsilantis and S\'{e}guinot \cite{t+j},
except that we adopt a numerical method to find the solution to the
equation for the photon direction, instead of simplifying it to a
$4^{th}$ order polynomial, which would allow an analytical
solution, but at the expense of introducing an additional source
of error. Furthermore, using our numerical method, we calculate
derivatives of the Cherenkov angle with respect to the measured
quantities which allows us to propagate the detector errors and
the chromatic dispersion to obtain an expected Cherenkov photon
resolution for each photon independently ($\sigma_\theta$). This
is useful, since the Cherenkov angle resolution varies
significantly even within one Cherenkov image. We use these
estimated errors when calculating particle ID likelihoods and use
them as to weight each photon when considering the average
Cherenkov angle for a track.

\subsection{Performance on Bhabha Events}

\begin{figure} [htb]
\vspace{-.28cm}
\centerline{\hspace{2.75 cm}\epsfxsize
1.24in\epsffile{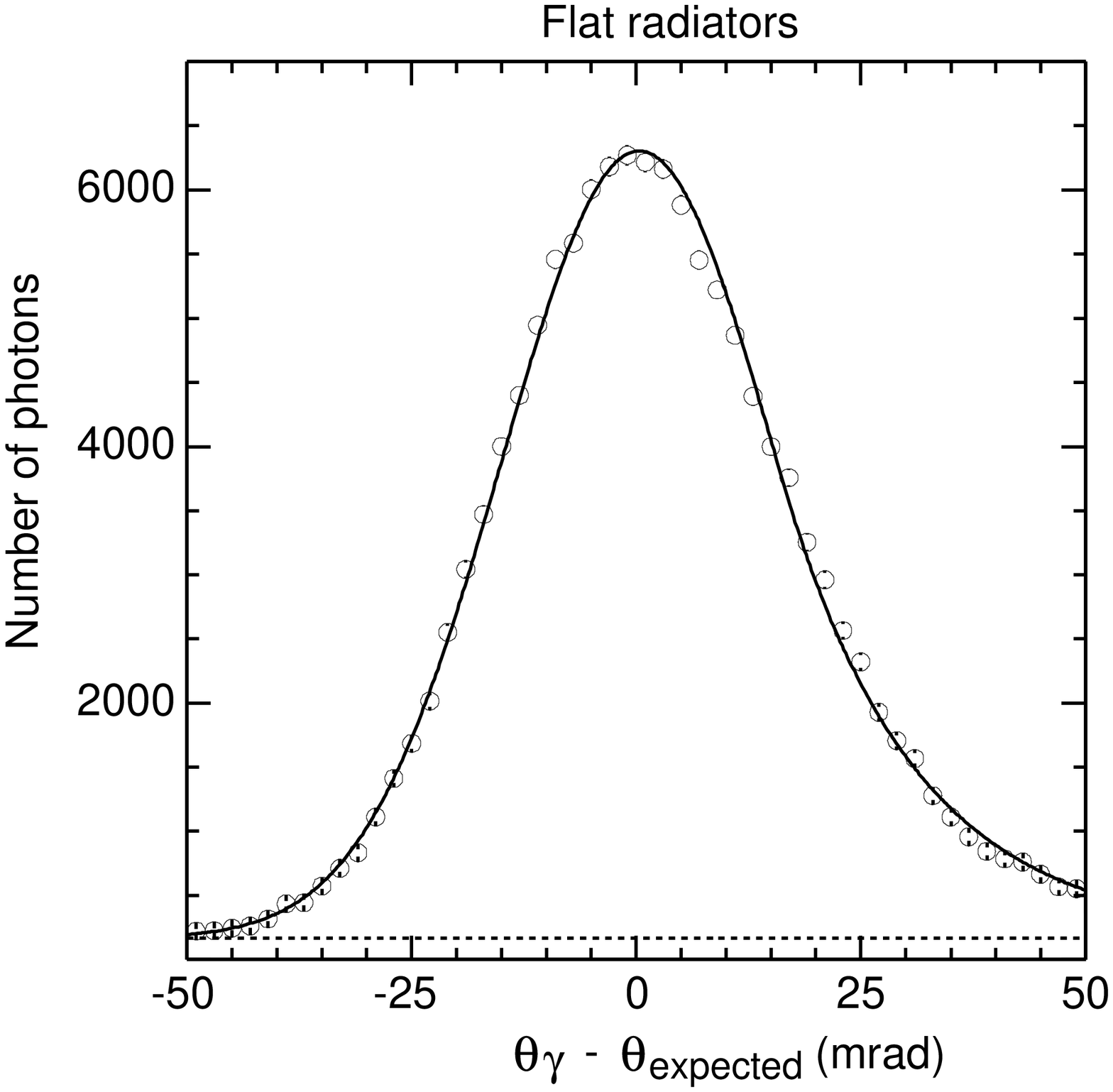}}
\centerline{\epsfxsize
2.48in\epsffile{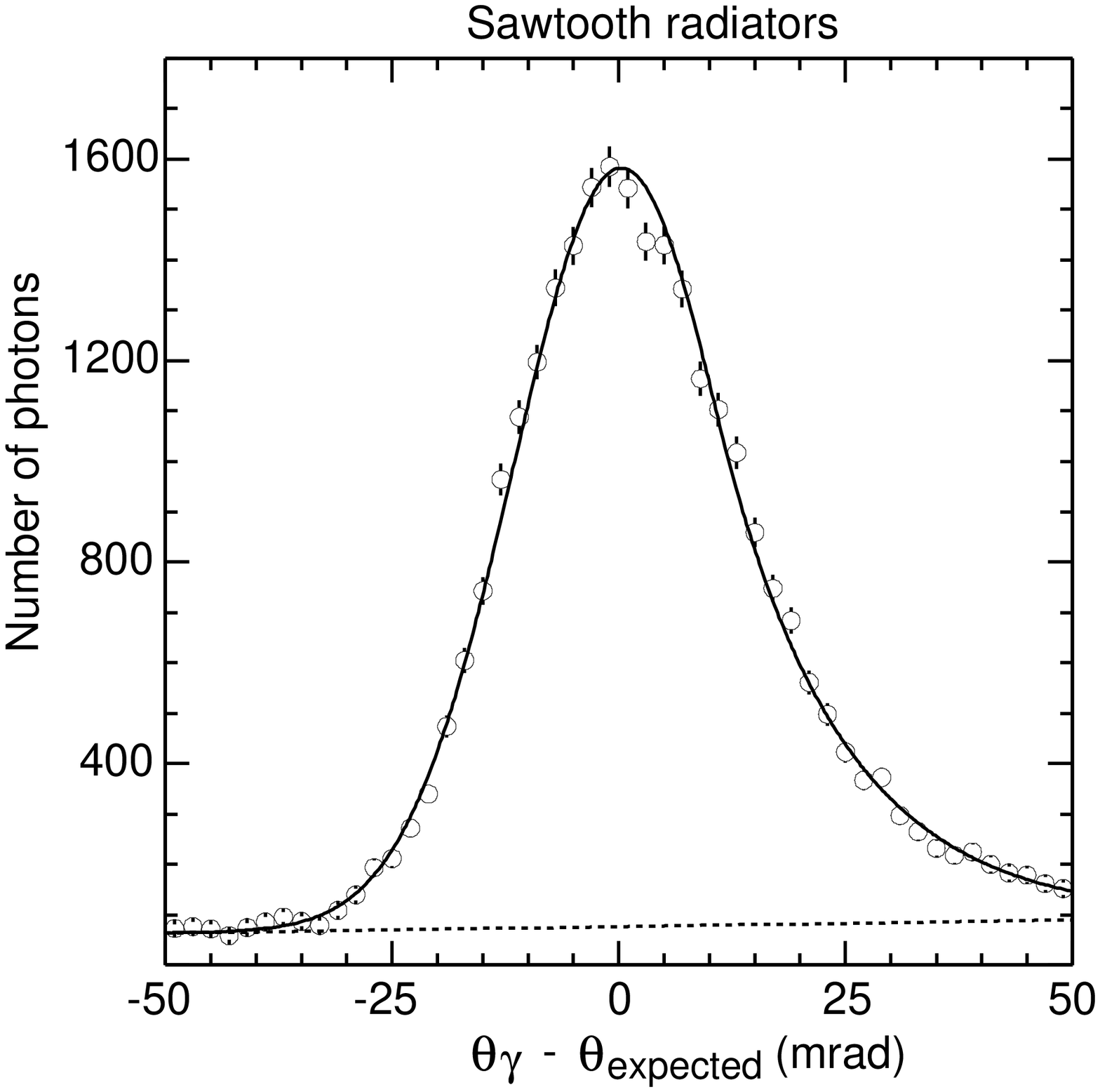}} \vspace{-.7cm}
\caption{\label{single_photon} The measured minus expected
      Cherenkov angle for each photon detected in Bhabha events,
      (top) for plane radiators and (bottom) for sawtooth radiators.
      The curves are fits to special line shape function (see text),
      while the lines are fits to a background polynomial.}
\vspace{-0.8cm}\end{figure}

We first view the physics performance on the simplest type of
events, Bhabha events and then subsequently in hadronic events.
The Cherenkov angle measured for each photon is shown in
Fig.~\ref{single_photon}.

We note that Bhabha events are very low multiplicity compared with
our normal hadronic events. They have two charged tracks present
while the hadronic events have an average charged multiplicity of
approximately 10. In addition, the hadronic events have on the
average 10 photons, mainly from $\pi^o$ decays. All of these
particles can interact in the calorimeter and the splash-back can
hit the RICH photon detector.

This Cherenkov angle spectrum for single photons has an asymmetric tail and modest
background. It is fit with a line-shape similar to that used by
for extracting photon signals from electromagnetic calorimeters
\cite{CBL}. The functional form is
\begin{equation}
P(\theta|\theta_{exp},\sigma_{\theta},\alpha,n)=
\end{equation}
\vspace{-2mm}
\begin{eqnarray*}
&A\cdot{\rm exp}\left[-{1\over 2}\left({{\theta_{exp}-\theta}\over \sigma_{\theta}}
\right)^2\right]~{{\rm for}~\theta<\theta_{exp}-\alpha\cdot\sigma_{\theta}}\\
&A\cdot{{\left({n\over \alpha}\right)^n e^{-{1\over 2}\alpha^2}
\over \left({{\theta_{exp}-\theta}\over \sigma_{\theta}}+{n\over \alpha}-\alpha\right)^n}}
~~~~~~~~~{{\rm for}~\theta>\theta_{exp}-\alpha\cdot\sigma_{\theta}},\\
&A^{-1}\equiv \sigma_{\theta}
\left[{n\over \alpha}{1\over {n-1}}e^{-{1\over 2}\alpha^2}
+\sqrt{\pi\over 2}\left(1+{\rm erf}\left({\alpha\over\sqrt{2}}\right)
\right)\right].
\end{eqnarray*}

Here $\theta$ is the measured angle, $\theta_{exp}$ is the ``true'' (or most
likely) angle and $\sigma_{\theta}$ is the angular resolution.
To use this formula, the parameter $n$ is fixed to value of about 5.

\begin{figure} [htb]
\vspace{-1.1cm}
\centerline{\epsfxsize 2.6in
\epsffile{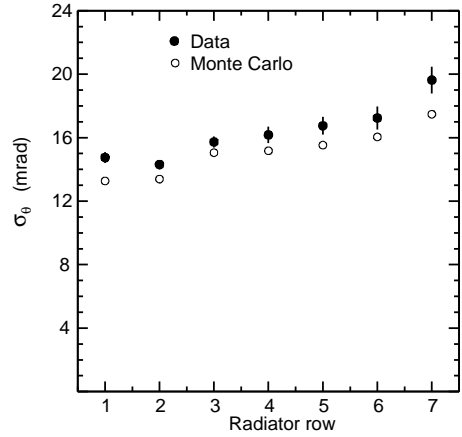}} \vspace{-0.7cm}
\caption{\label{bhares_ang} The values of the angular resolution
for single photons for data compared with Monte Carlo simulation
as a function of radiator ring. Sawtooth radiators are in rings 1
and 2, near the center of the detector.}
\vspace{-0.7cm}\end{figure}

The data in Fig.~\ref{single_photon} are fit using this signal
shape plus a polynomial background function. We compare the
results of these fits for the resolution parameter
$\sigma_{\theta}$ as a function of radiator ring for data and
Monte Carlo simulation in Fig.~\ref{bhares_ang}. The single photon
resolution averaged over the detector solid angles are 14.7 mr for
the flat radiator and 12.2 mr for the sawtooth.

\begin{figure} [htb]
\vspace{-.9cm} \centerline{\epsfxsize 2.5in
\epsffile{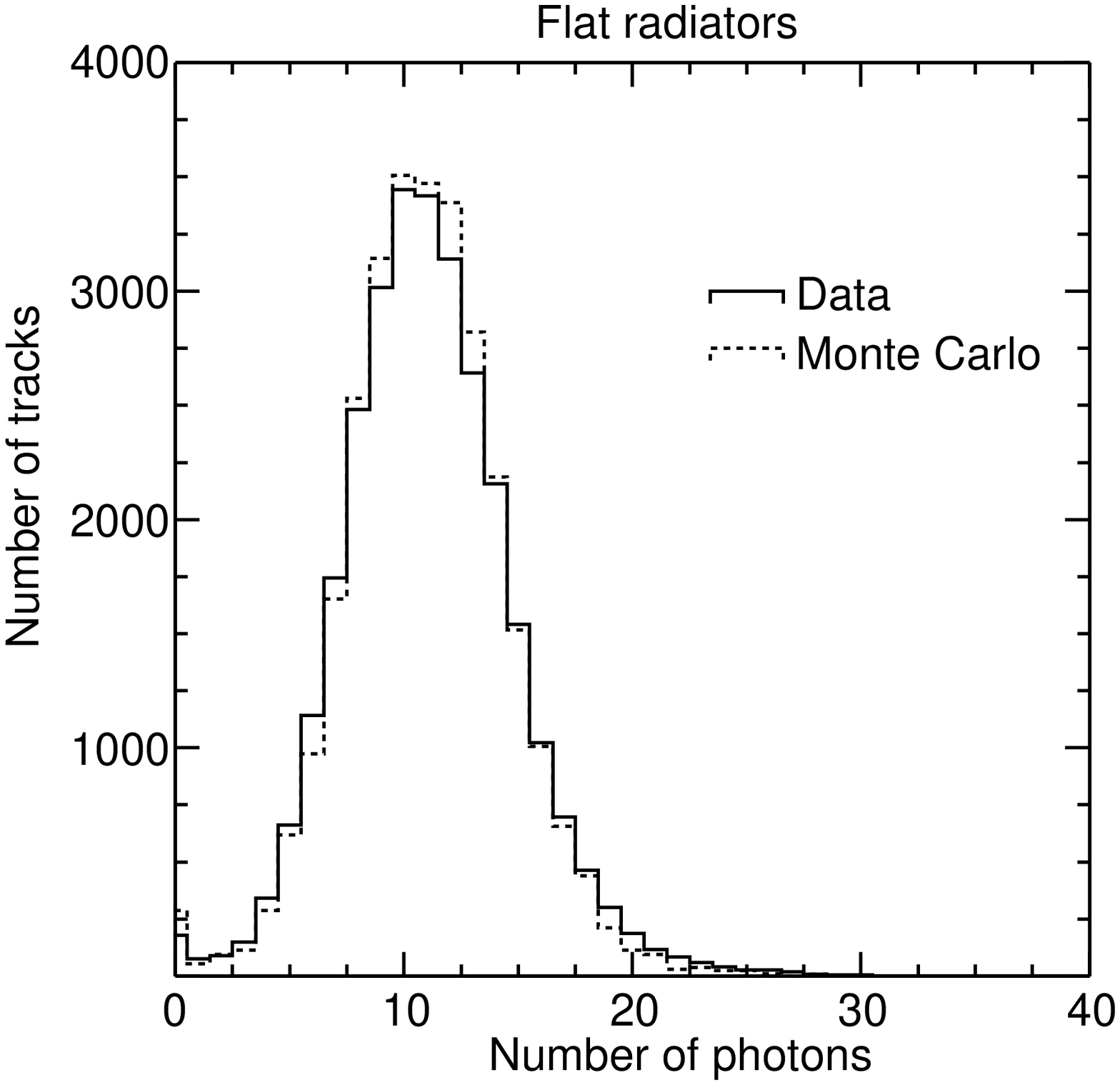}} \vspace{-0.7cm}
\centerline{\epsfxsize 2.5in \epsffile{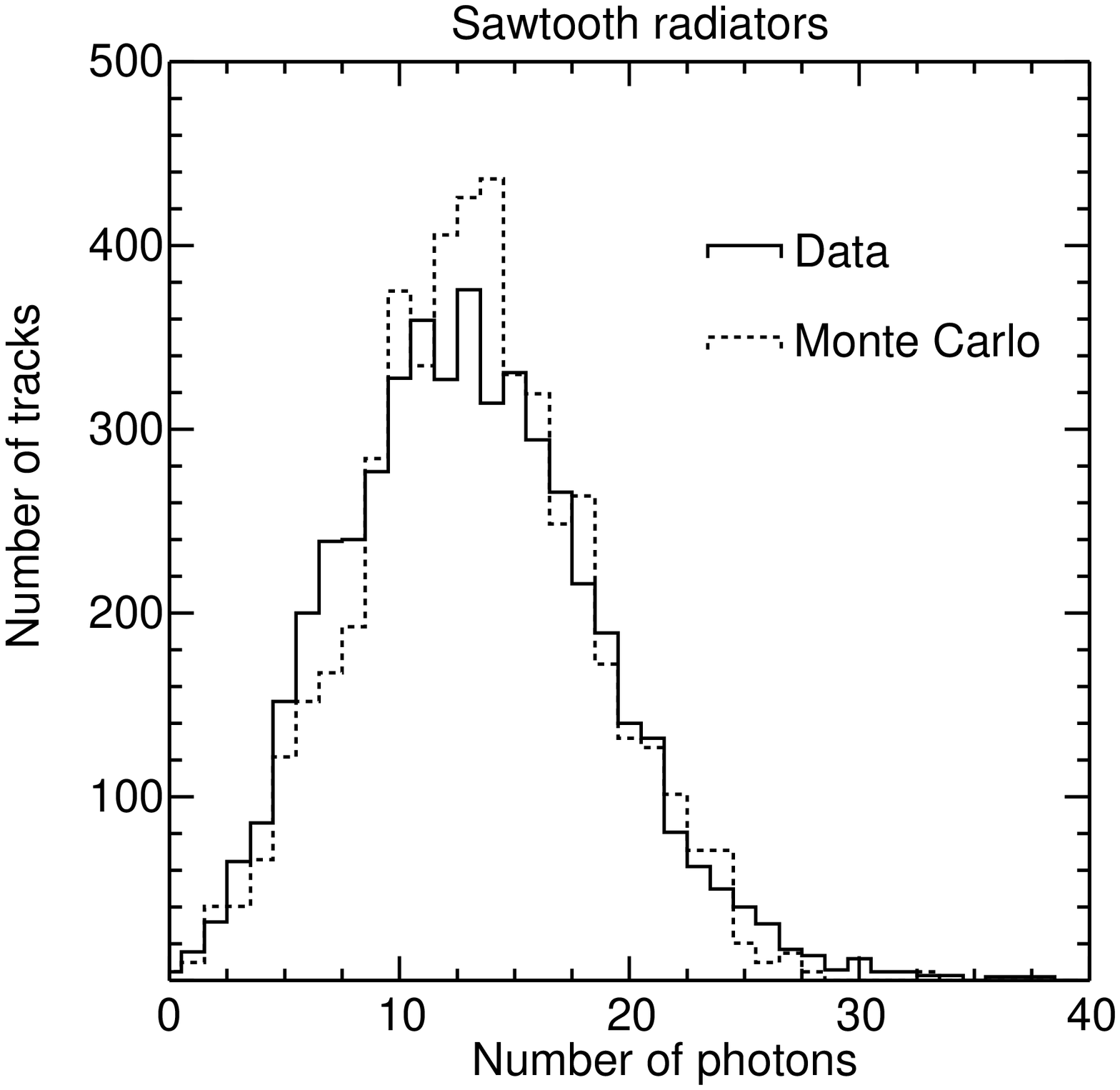}}
\vspace{-.7cm} \caption{\label{ngamma} The number of photons
detected on Bhabha tracks (top) for plane radiators and (bottom)
for sawtooth radiators. The dashed lines are predictions of the
Monte Carlo simulation.} \vspace{-0.7cm}\end{figure}

\begin{figure} [htb]
\centerline{\hspace{.9in}\epsfxsize 1.3in
\epsffile{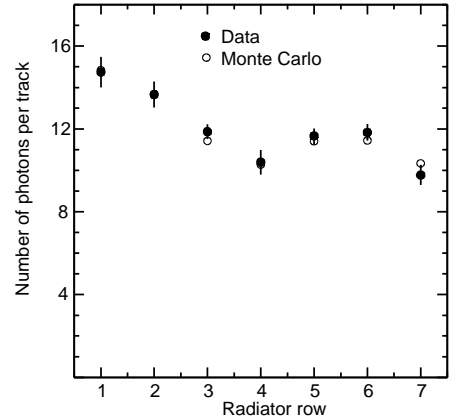}}
\caption{\label{bha_photon} The number of photons as a function of
radiator ring.} \vspace{-0.7cm}\end{figure}

The number of photons per track within a $\pm 3\sigma$ of the
expected Cherenkov angle for each photon is shown in
Fig.~\ref{ngamma} and shown as a function of radiator ring in
Fig.~\ref{bha_photon}. Averaged over the detector, and subtracting
the background we have a mean number of 10.6 photons with the flat
radiators and 11.9 using the sawtooth radiators.

\begin{figure} [htb]
\vspace{1cm} \centerline{\hspace{1in}\epsfxsize .922in
\epsffile{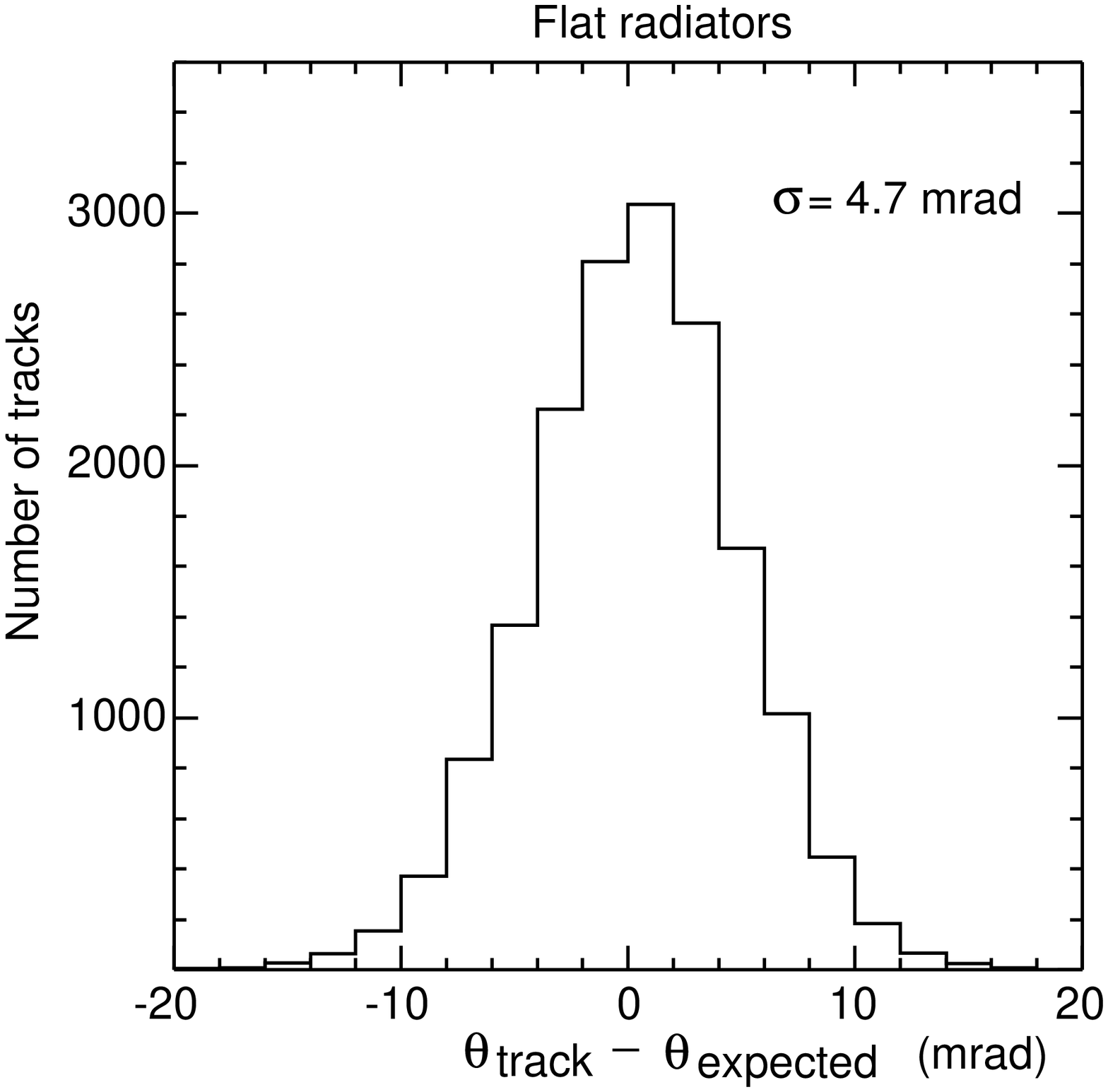}} \vspace{-0.4cm}
\centerline{\epsfxsize 2.65in\epsffile{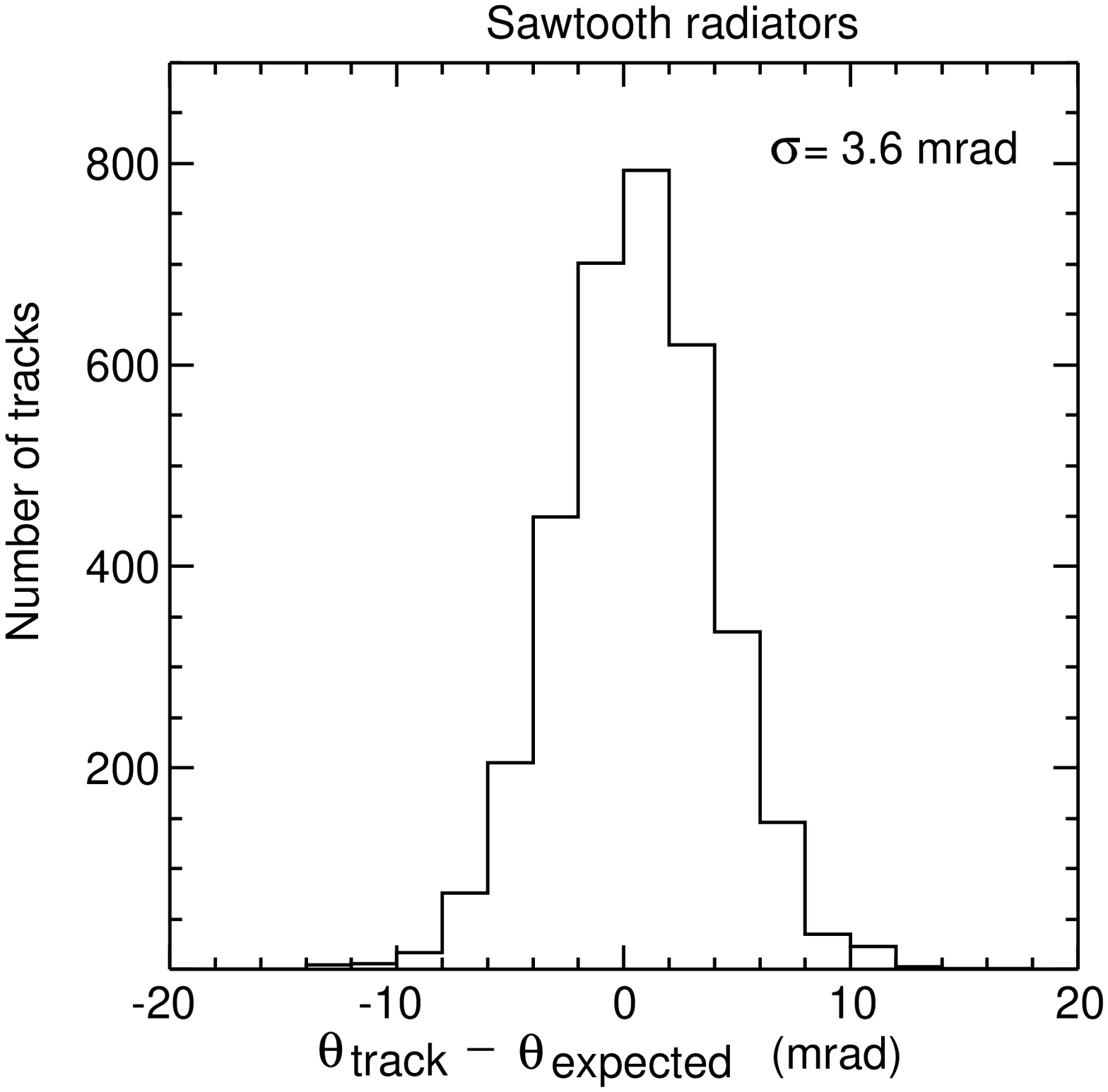}}
\vspace{-.7cm} \caption{\label{trk_res} Track resolutions in
Bhabha events,
   (left) for plane radiators and (right) for sawtooth radiators.}
\vspace{-0.7cm}\end{figure}

The resolution per track is obtained by taking a slice within $\pm 3\sigma$
of the expected Cherenkov angle for each photon and forming an
average weighted by $1/\sigma^2_{\theta}$.
These track angles are shown in Fig.~\ref{trk_res}.

\begin{figure} [htb]
\centerline{\hspace{.5in}\epsfxsize 1.3in
\epsffile{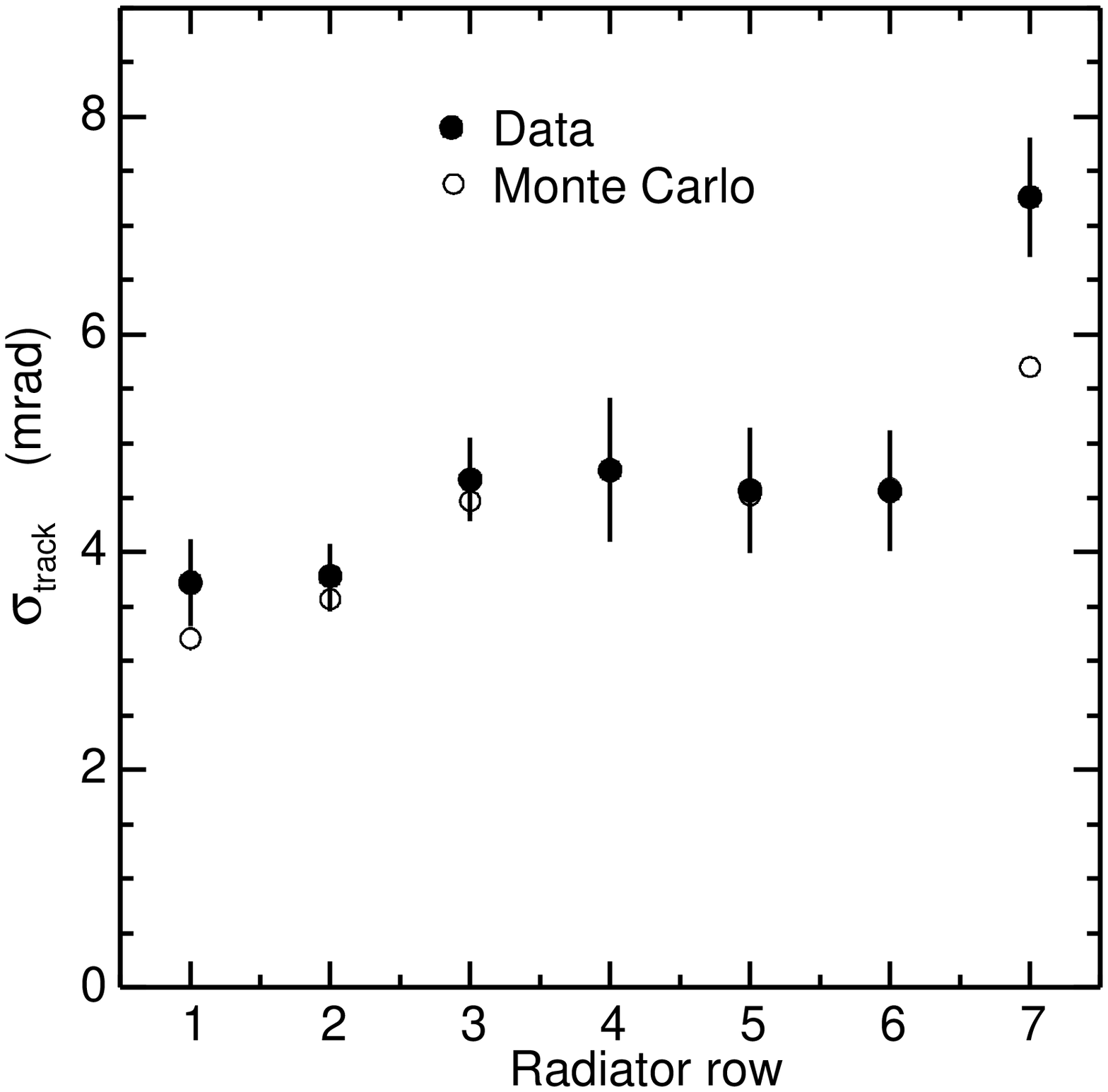}} \vspace{-.4cm}
\caption{\label{bha_track} Cherenkov angle resolutions per track
as a function of radiator ring for Bhabha events.}
\vspace{-0.7cm}\end{figure}

The rms spreads of these distributions are identified as the track
resolutions. We obtain 4.7 mr for the flat radiators and 3.6 mr
for the sawtooth. The resolutions as a function of radiator ring
are shown in Fig.~\ref{bha_track}.

\begin{figure} [htb]
\centerline{\hspace{.4in}\epsfxsize 2.1in
\epsffile{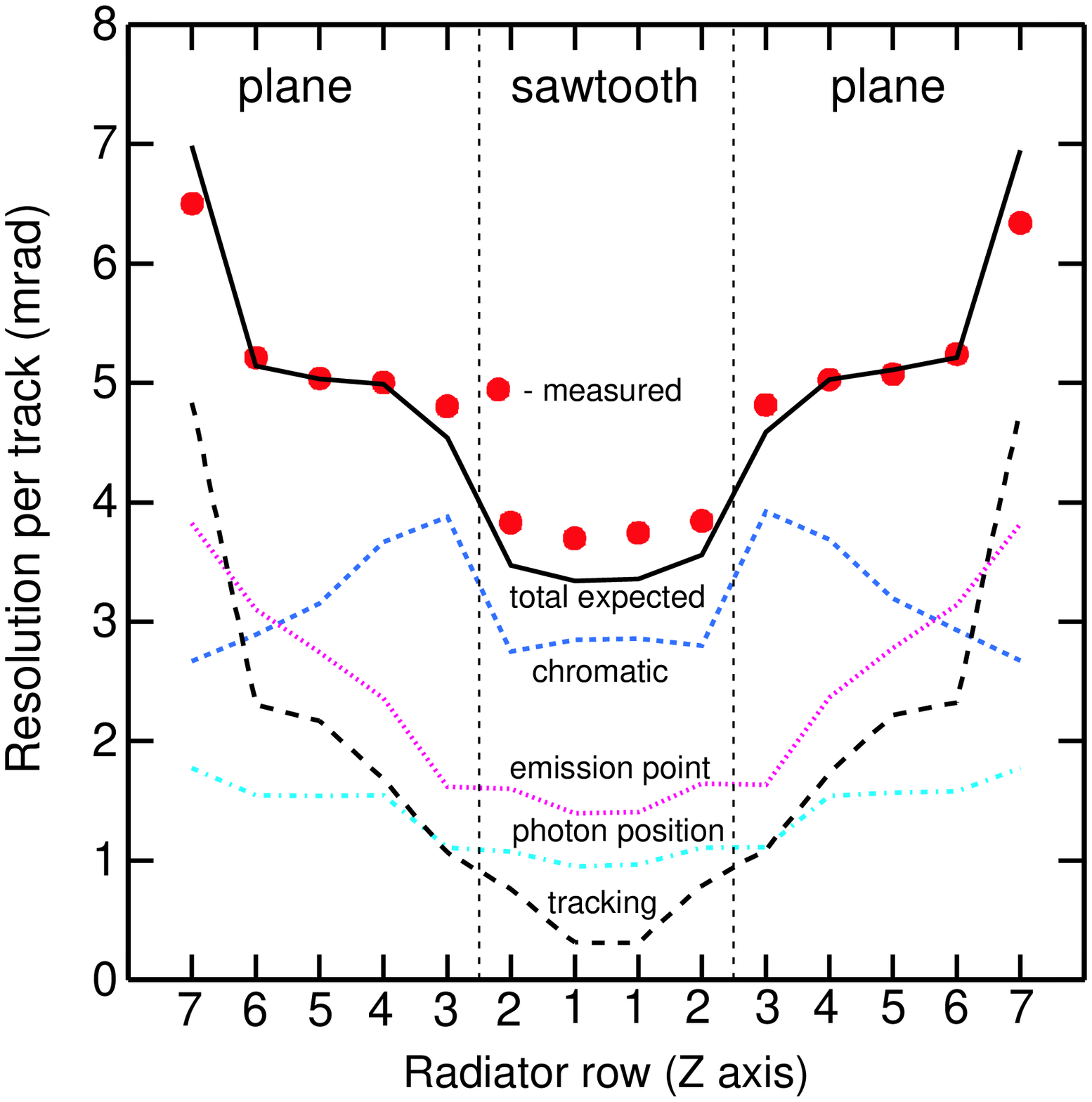}} \vspace{-.4cm}
\caption{\label{res} Different components of the Cherenkov angle
resolutions per track as a function of radiator ring for Bhabha
events. The points are the data and the solid line is the sum of
the predicted resolution from each of the components indicated on
the figure.} \vspace{-0.7cm}\end{figure}

The Cherenkov angular resolution is comprised of several different
components. This include error on the location of the photon emission point,
the chromatic dispersion, the position error in the reconstruction of the
detected photons and finally the error on determining the charged track's direction and position.
These components are compared with the data in Fig.~\ref{res}.

\subsection{Performance on Hadronic Events}

To resolve overlaps between Cherenkov images for different tracks
we find the most likely mass hypotheses. Photons that match the
most hypothesis within $\pm3\sigma$ are then removed from
consideration for the other tracks. To study the RICH performance
in hadronic events we use  $D^{*+}\to\pi^+ D^0$, $D^0\to K^-\pi^+$
events. The charge of the slow pion in the $D^{*+}$ decay is
opposite to the kaon charge in subsequent $D^0$ decay. Therefore,
the kaon and pion in the $D^0$ decay can be identified without use
of the RICH detector. The effect of the small combinatorial
background is eliminated by fitting the $D^0$ mass peak in the
$K^+\pi^-$ mass distribution to obtain the number of signal events
for each momentum bin.

\begin{figure} [htb]
\centerline{\hspace{.5in}\epsfxsize
1.5in\epsffile{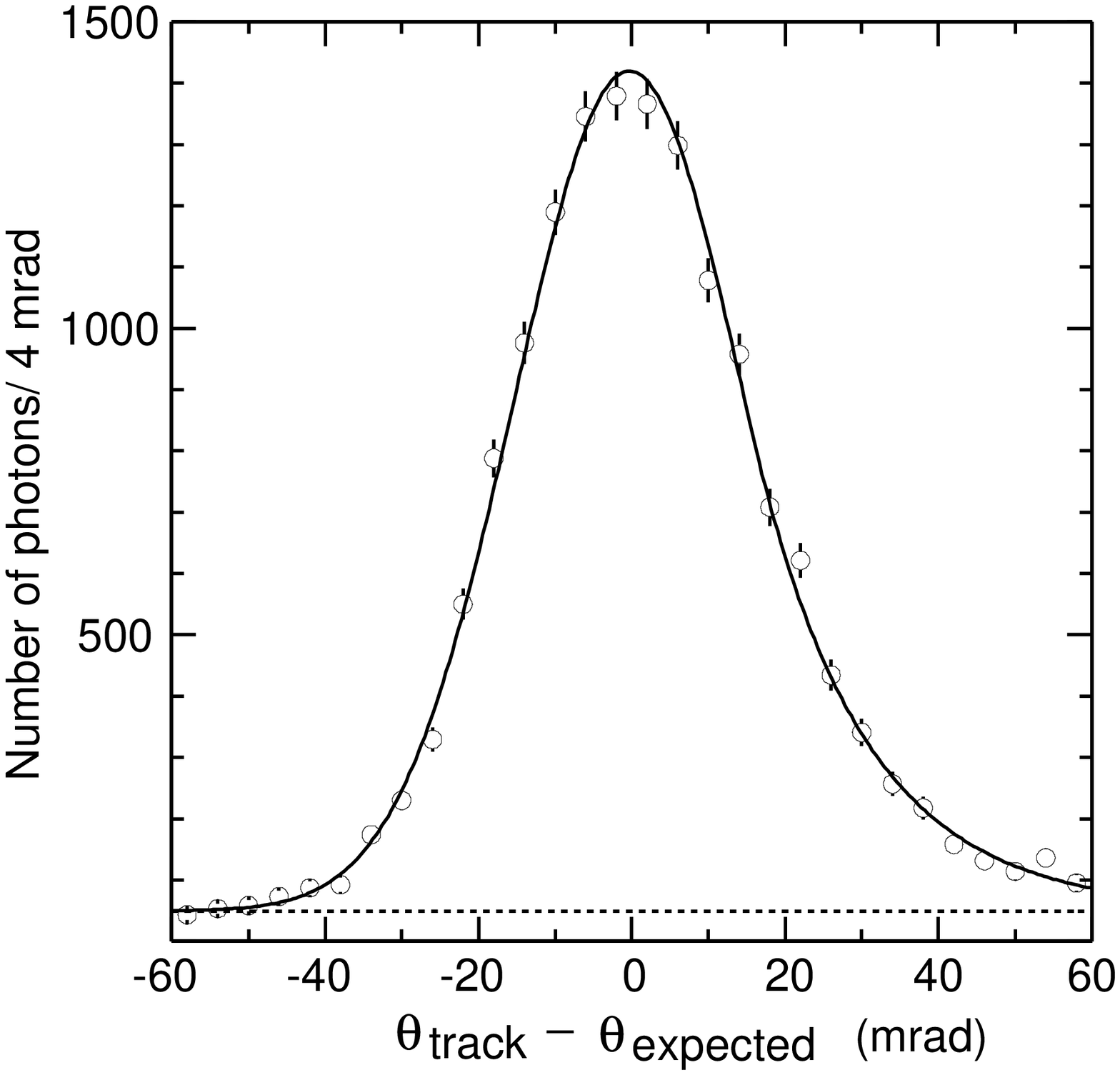}}
\centerline{\hspace{.4in}\epsfxsize
1.6in\epsffile{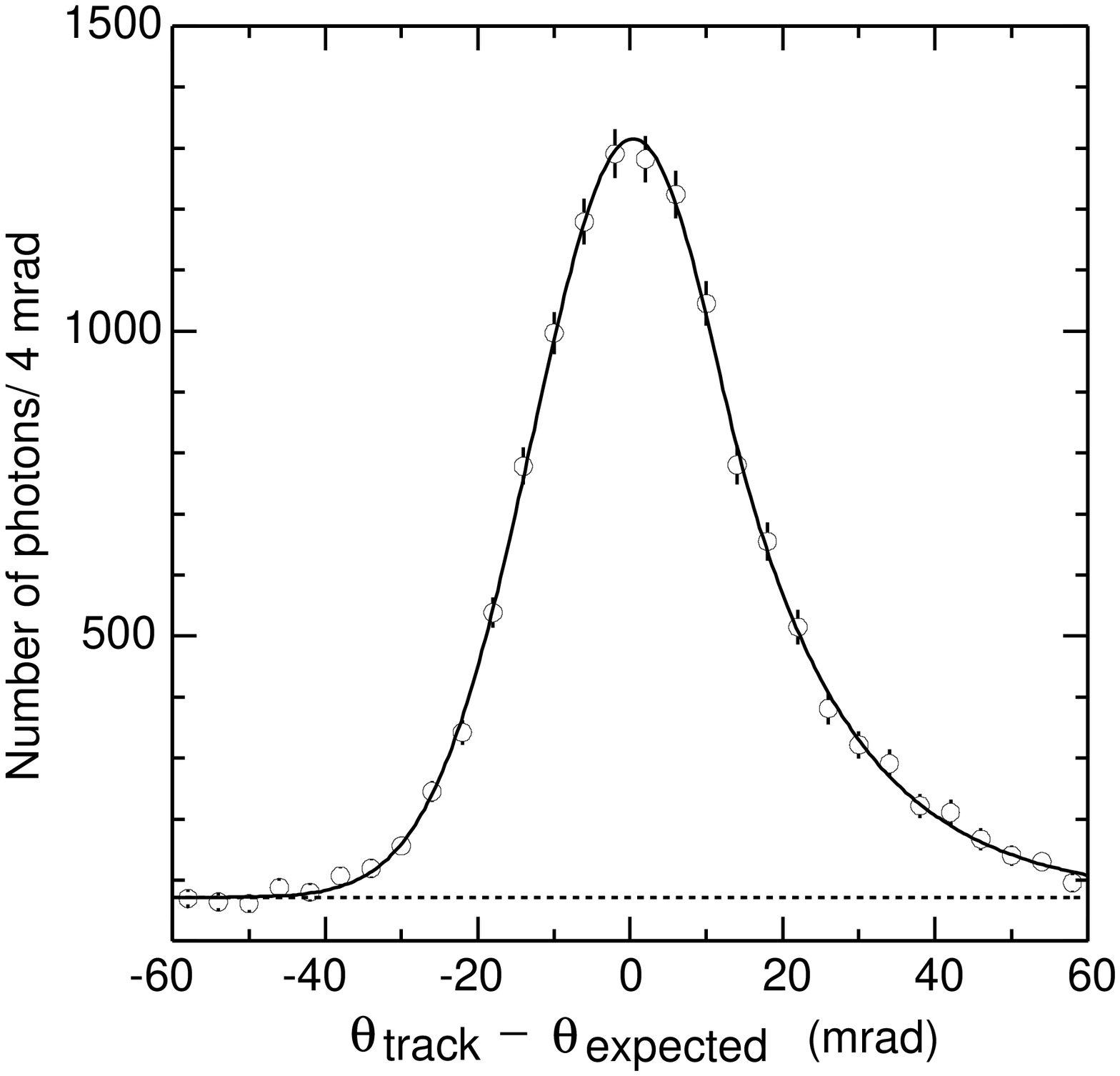}}
\vspace{-.7cm} \caption{\label{single_photon_hadrons} The measured
minus expected
      Cherenkov angle for each photon detected in hadronic events,
      (top) for plane radiators and (bottom) for sawtooth radiators.
      The curves are fits to special line shape function (see text),
      while the lines are fits to a background polynomial.}
\vspace{-0.7cm}\end{figure}

Single-photon Cherenkov angle distributions
obtained on such identified
kaons with the momentum above 0.7 GeV/c are plotted
in Fig.~\ref{single_photon_hadrons}.
Averaged over all radiators, the single-photon resolution is
13.2 mr and 15.1 mr for sawtooth and flat radiators
respectively.
The background fraction within $\pm3\sigma$ of the
expected value is 12.8\% and 8.4\%.
The background-subtracted mean photon yield is 11.8 and 9.6.
Finally the per-track Cherenkov angle resolution is
3.7 mr and 4.9 mr.

\subsection{Particle ID Likelihoods}

For parts of the Cherenkov image for the sawtooth radiator, and for
tracks intersecting more than one radiator there are some
optical path ambiguities that impact the Cherenkov angle calculations.
In the previous section we bypassed this problem by selecting the optical
path that produces the closest Cherenkov angle to the expected one
($\theta^h_{exp}$) for the given particle hypothesis ($h$).
There is some loss of information in this procedure, therefore,
we use the likelihood method to perform particle identification
instead of the per-track average angle.
The likelihood method weights each possible optical path
by the optical probability ($P_{opt}$),
which includes length of the radiation path
and the refraction probabilities obtained by the inverse ray tracing method:
\begin{eqnarray*}
L_h = \prod_{j=1}^{No.\,of\,\gamma s} \left\{
P_{{background}} +  \hbox{\qquad\qquad} \phantom{\sum_{opt} P_{opt}^j} \right. \\
   \left. \sum_{opt} P_{opt}^j
    \cdot P_{{signal}}\left( \theta_\gamma^{opt,\,j} |
         \theta_{exp}^h,\,\sigma_\theta^{opt,\,j} \right) \right\}
\end{eqnarray*}
where, $L_h$ is the likelihood for the particle hypothesis $h$
($e$, $\mu$, $\pi$, $K$ or $p$), $P_{{background}}$
is the background probability approximated by a constant
and $P_{{signal}}$ is the signal probability
given by the line-shape defined previously.
In principle, the likelihood could include all hits
in the detector. In practice, there is no point in inspecting hits
which are far away from the regions where photons are expected for
at least one of the considered hypotheses (we use $\pm5\sigma$ cut-off).

An arbitrary scale factor in the likelihood definition cancels
when we consider likelihood ratios for two different hypotheses.
The likelihood conveniently folds in information about values of
the Cherenkov angles and the photon yield for each hypothesis. For
well separated hypotheses (typically at lower momenta) it is the
photon yield that provides the discrimination. For hypotheses that
produce Cherenkov images in the same area of the detector, the
values of the Cherenkov angles do the job. Since our likelihood
definition does not know about the radiation momentum threshold,
the likelihood ratio method can be only used when both hypotheses
are sufficiently above the thresholds. When one hypothesis is
below the radiation threshold we use a value of the likelihood for
the hypothesis above the threshold to perform the discrimination.

\begin{figure} [htb]
\vspace{-1.3cm} \centerline{\epsfxsize
3.0in\epsffile{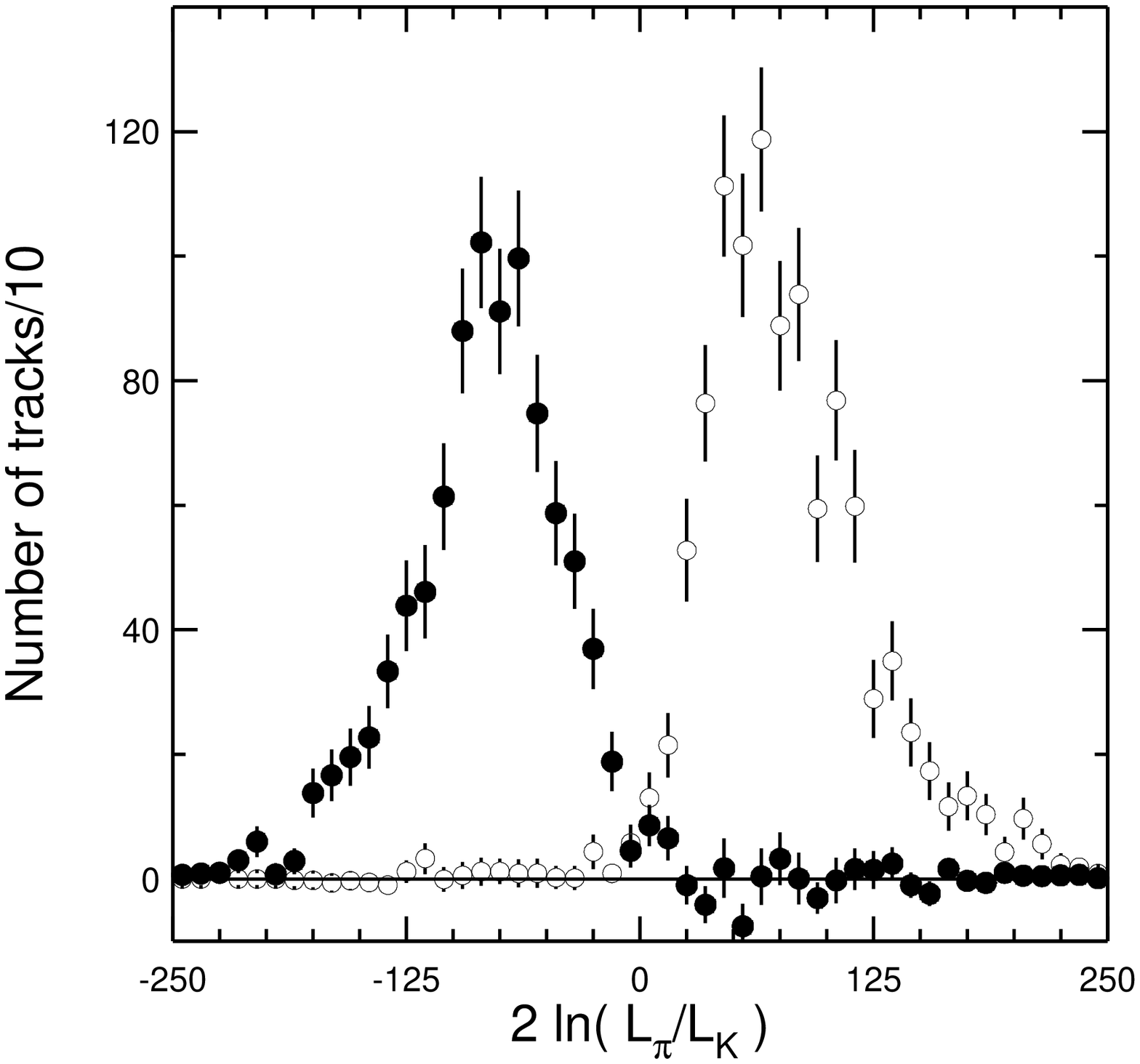}} \vspace{-.7cm}
\caption{\label{likelihood_ratio}
      Distribution of $2\ln\left(L_\pi/L_K\right)$ $\sim\chi^2_K-\chi^2_\pi$
      for 1.0-1.5 GeV/c kaons (filled)
      and pions (open) identified with the $D^*$ method.}
\vspace{-0.7cm}\end{figure}
\begin{figure} [hbt]
\vspace{.2cm} \centerline{\epsfxsize
2.8in\epsffile{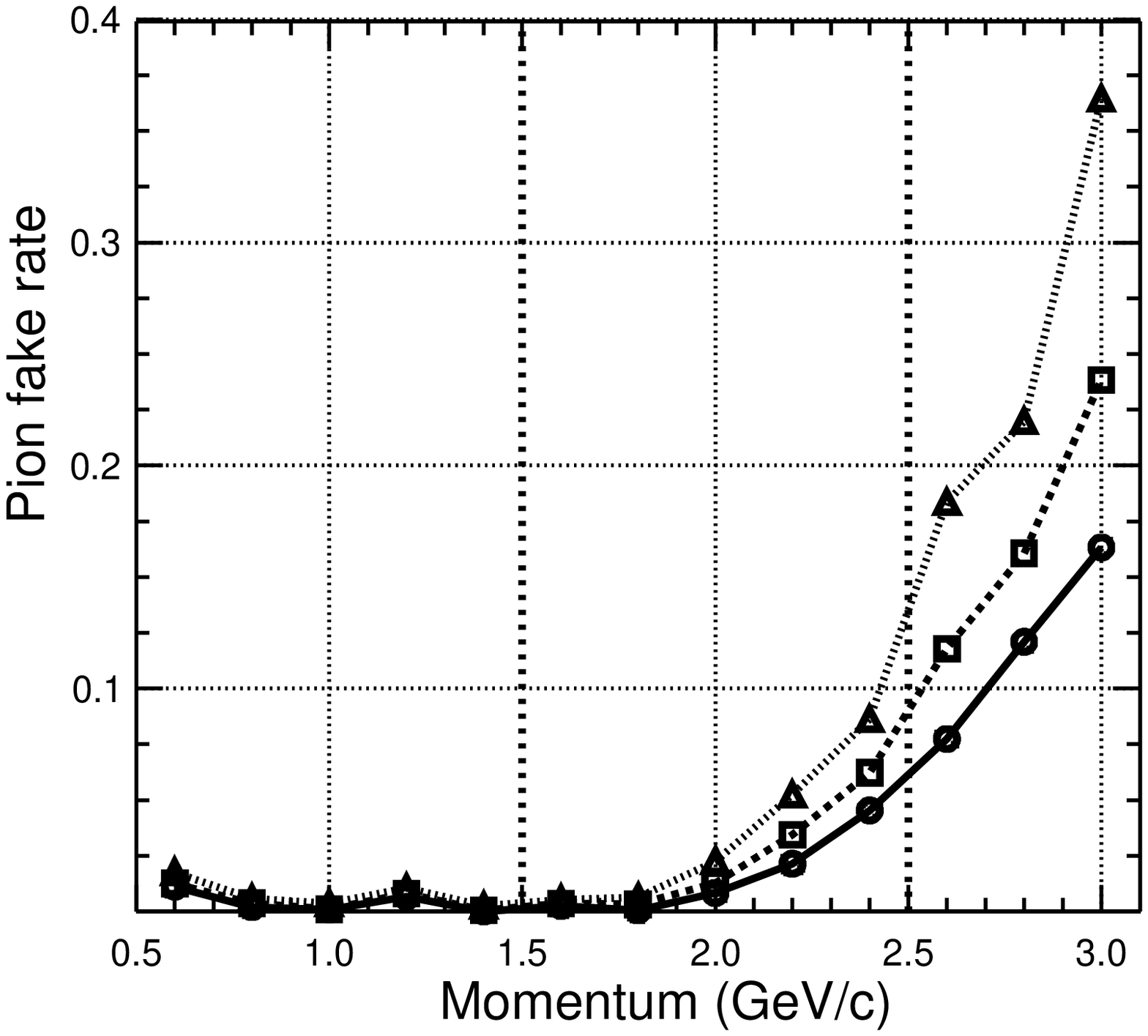}} \vspace{-.7cm}
\caption{\label{id_vs_momentum}
      Pion fake rate as a function of particle momentum for
      kaon efficiency of 80\%\ (circles), 85\%\ (squares) and
      90\%\ (triangles).} \vspace{-0.7cm}\end{figure}

The distribution of the $2\ln\left(L_\pi/L_K\right)$, is expected
to behave as the difference $\chi^2_K-\chi^2_\pi$. This $\chi^2$
difference obtained for 1.0-1.5 GeV/c kaons and pions identified
with the $D^*$ method is plotted in Fig.~\ref{likelihood_ratio}.
Cuts at different values of this variable produce identification
with different efficiency and fake rate. Pion fake rate for
different values of kaon identification efficiency is plotted as a
function of particle momentum in Fig.~\ref{id_vs_momentum}.

\subsection{Conclusions}
We have successfully constructed and operated a large, complex
RICH detector in a particle physics experiment for over three
years. About 98\% of the detector is operational (1\% loss due to
one broken wire and 1\% due to electronic failures).

The particle momenta for $B$ meson decay products seen by CLEO are
less than $2.65$ GeV/c. The detector provides excellent separation
between pions and kaons at and below this cutoff. Separation
between kaons and protons extends to even higher momentum, where
it is used in charm baryon studies. Thus, the physics performance
has met design criteria.

CLEO currently is making an extensive study of Upsilon decays and
proposes to study decays of charm mesons and charmonium decays
(called CLEO-c \cite{CLEO-c}). For these measurements the beam
energy will be lowered and the maximum particle momenta will be
about $1.5$ GeV/c. At these momentum the particle identification
fake rates are at the 1\% level.

\subsection{Acknowledgments}
This work was supported by the U. S. National Science Foundation
and Department of Energy. We thank Tom Ypsilantis and Jacques
S\'{e}guinot for suggesting the basic technique. We thank the
accelerator group at CESR for excellent efforts in supplying
luminosity.


\newpage
\begin{thebibliography}{99}

\bibitem{CLEOII}
 Y. Kubota \etal , Nucl. Instr. Meth. A320 (1992) 66.
\bibitem{Artu98}
 M.~Artuso, ``Progress Towards CLEO III'',
 in the Proceedings of the
 XXIX International Conference on High Energy Physics, Vancouver,
 Ed. by A. Astbury et al., World Scientific, Singapore,
 vol. 2, p 1552, [hep-ex/9811031] (1998).
\bibitem{Kopp96}
 S.E.~Kopp, Nucl.\ Instr.\ Meth.\ A384 (1996) 61.

\bibitem{Arno92}
A similar system was tested previously, see
R.~Arnold et al., Nucl.\ Instr.\ Meth.\ A314 (1992) 465;
J.-L.~Guyonnet et al., Nucl.\ Instr.\ Meth.\ A343 (1994) 178;
J.~S\'{e}guinot et al., Nucl.\ Instr.\ Meth.\ A350 (1994) 430.
\bibitem{testbeam}
M.~Artuso et al., Nucl.\ Instr.\ Meth.\ A441 (2000) 374-392
\bibitem{t+j}
 T.~Ypsilantis and J.~S\'{e}guinot, Nucl.\ Instr.\ Meth.\ A343 (1994)
30.
\bibitem{efimov}
 A. Efimov and S. Stone, { Nucl. Instr. and Meth.} { A371} (1996) 79.

\bibitem{expon}
 R. Bouclier et al., Nucl.\ Instr.\ Meth.\ A205 (1983) 205.

\bibitem{DSP}
This alorithim is executed by a DSP located on the data boards
before the data are sparsified.

\bibitem{Peterson}
 D. Peterson et al., Nucl.\ Instr.\ Meth.\ A478 (2002) 142.

 \bibitem{CBL}
T. Skwarnicki, ``A Study of the Radiative Cascade Transitions Between the
Upsilon-Prime and Upsilon Resonances," DESY F31-86-02 (thesis, unpublished)
(1986).

\bibitem{CLEO-c}
I. Shipsey, ``CLEO-c and CESR-c: Allowing Quark Flavor Physics to
Reach its Full Potential," to appear in the proceedings of Flavor
Physics and CP Violation (FPCP) May, 2002. Univ. of Pennsylvania,
Philadelphia, PA, [hep-ex/0207091] (2002).

\end{thebibliography}
\end{document}